\documentclass[preprint,proceedings]{rmaa}

\SetYear{2002} \SetConfTitle{Galaxy Evolution: Theory and Observations}

\title{Joint Formation of QSOs and Spheroids}
\author{G.L. Granato\altaffilmark{1}, L. Silva\altaffilmark{2}, 
  G. De Zotti\altaffilmark{1}, and L. Danese\altaffilmark{3}}
\altaffiltext{1}{INAF-Osservatorio Astronomico di Padova, Italy.}
\altaffiltext{2}{INAF-Osservatorio Astronomico di Trieste, Italy.}
\altaffiltext{3}{SISSA, Trieste, Italy.}
\shortauthor{Granato et al.}
\shorttitle{Joint formation QSOs-Spheroids}
\fulladdresses{\item G.L. Granato and G. De Zotti: INAF-Osservatorio Astronomico di Padova,
Via Osservatorio 5, I35122, Padova, Italy (granato@pd.astro.it, dezotti@pd.astro.it).
\item L. Silva: INAF-Osservatorio Astronomico di Trieste, 
Via Tiepolo 11, I34131, Trieste, Italy (silva@ts.astro.it).
\item L. Danese: SISSA, Via Beirut 4, I34131, Trieste, Italy (danese@sissa.it).}

\listofauthors{G.L. Granato, L. Silva, G. De Zotti, \& L. Danese}
\indexauthor{Granato, G. L.} \indexauthor{Silva, L.} \indexauthor{De Zotti, G.} 
\indexauthor{Danese, L.}

\abstract{In view of the extensive evidence of tight inter-relationships
between spheroidal galaxies (and galactic bulges) with massive
black holes hosted at their centers, a consistent model must deal
jointly with the evolution of the two components. We describe one
such model, which successfully accounts for the local luminosity
function of spheroidal galaxies, for their photometric and
chemical properties, for deep galaxy counts in different
wavebands, including those in the (sub)-mm region which proved to
be critical for current semi-analytic models stemming from the
standard hierarchical clustering picture, for clustering
properties of SCUBA galaxies, of EROs, and of LBGs, as well as for
the local mass function of massive black holes and for quasar
evolution. Predictions that can be tested by surveys carried out
by SIRTF are presented.}

\addkeyword{dust,extinction} 
\addkeyword{galaxies: formation}
\addkeyword{quasars: general} \addkeyword{Infrared: galaxies} 
\addkeyword{Cosmology: theory}

\begin{document}
\maketitle

\section{Introduction}
\label{sec:intro}
The hierarchical clustering model with a scale invariant spectrum
of density perturbations in a Cold Dark Matter (CDM) dominated
universe has proven to be remarkably successful in matching the
observed large-scale structure as well as a broad variety of
properties of galaxies of the different morphological types
(Granato et al. 2000 and references therein). However, serious
shortcomings of this scenario have also become evident in recent
years.

\begin{figure*}                                                 
  \includegraphics[width=\textwidth,height=7cm]{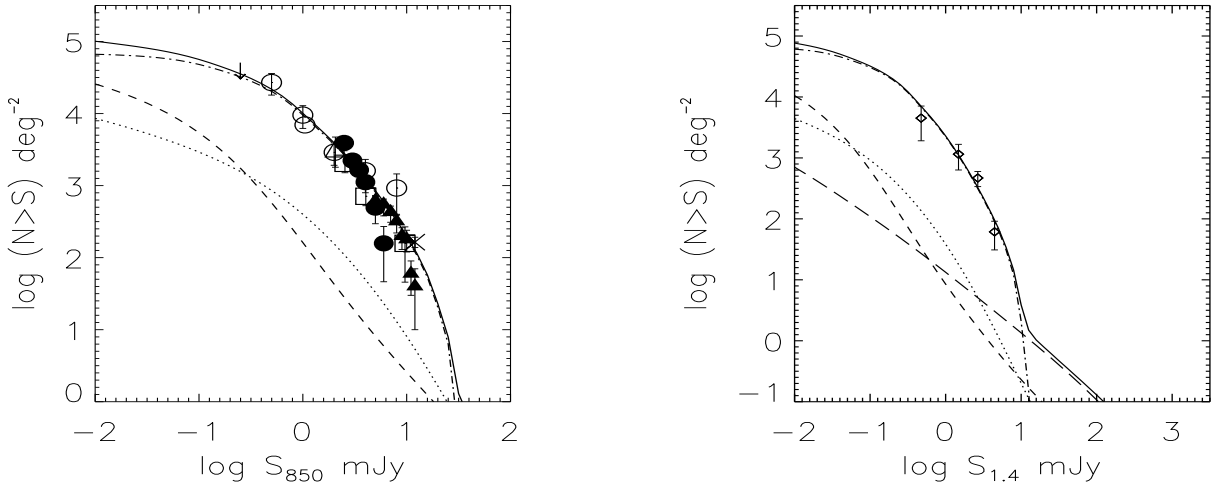} 
  \caption{Integral source counts at $850\,\mu$m and $1.4$ mm
  predicted by the model by Granato et al.\ (2001) compared 
  with observations. The dotted, dashed, dot-dashed lines show 
  the contributions of starburst, spiral, and forming elliptical galaxies respectively.
  The long-dashed (at 1.4 mm only) gives the counts of radio sources (Toffolatti et al.\ 1998).}  
  \label{fig1}                                                       
\end{figure*}

The critical point can be traced back to the relatively large amount of power on
small scales predicted by this model which would imply far more dwarf galaxies or
substructure clumps within galactic and cluster mass halos than are observed (the
so-called ``small-scale crisis"), unless star formation in small
objects is strongly suppressed (or the small scale power is reduced by modifying
the standard model). At the other extreme of the galaxy mass function 
we have another strong discrepancy with model predictions, 
that we might call ``the massive galaxy
crisis''. Even the best semi-analytic models hinging upon the standard
picture for structure formation in the framework of the hierarchical
clustering paradigm, are stubbornly unable to account for the (sub)-mm
(SCUBA and MAMBO, Figure~\ref{fig1}) counts of galaxies, 
most of which are
probably massive objects undergoing a very intense starburst (with star
formation rates $\sim 1000\,\hbox{M}_\odot\,\hbox{yr}^{-1}$) at $z>2$.
Recent optical data confirm that most massive
ellipticals were already in place and (almost) passively evolving up to
$z\simeq 1$--1.5. These data are more consistent with the traditional
``monolithic'' approach whereby giant ellipticals formed most of their
stars in a single gigantic starburst at substantial redshifts, and
underwent essentially passive evolution thereafter.
In the canonical hierarchical clustering paradigm the smallest objects
collapse first and most star formation occurs, at relatively low rates,
within relatively small proto-galaxies, that later merged to form larger
galaxies. Thus the expected number of galaxies with very intense star
formation is far less than detected in SCUBA and MAMBO surveys and the
surface density of massive evolved ellipticals at $z\gtrsim 1$ is also
smaller than observed. The ``monolithic'' approach, however, is
inadequate to the extent that it cannot be fitted in a consistent
scenario for structure formation from primordial density fluctuations.

\section{Relationships between quasar and galaxy evolution}
\label{sec:evo}
The above difficulties, affecting even the best current recipes,
may indicate that new ingredients need to be taken into account. A
key new ingredient may be the mutual feedback between formation
and evolution of spheroidal galaxies and of active nuclei residing
at their centers. In this framework, Granato et al. (2001)
elaborated the following scheme.

Feed-back effects, from supernova explosions and from active nuclei
(note that supernova feedback alone falls short of solving the dearth of dwarf
galaxies, but photo-ionization by the UV background re-ionizing the
inter-galactic medium (IGM) could do the job)
delay the collapse of baryons in smaller clumps while large ellipticals
form their stars as soon as their potential wells are in place; {\it the
canonical hierarchical CDM scheme -- small clumps collapse first -- is
therefore reversed for baryons}. Large spheroidal galaxies therefore
undergo a phase of high (sub)-mm luminosity.

At the same time, the central black-hole (BH) grows by accretion and the
quasar luminosity increases; when it reaches a high enough value, its
action (ionization and heating of the gas) 
stops the star formation and eventually expels the residual gas.
This explains the observed correlation between BH
and host spheroidal masses. The same mechanism distributes
in the IGM a substantial fraction of metals and may pre-heat the IGM. The
onset of quasar activity (and the corresponding end of star formation)
occurs earlier for more massive objects. 
The duration of the starburst increases with decreasing mass from
$\sim 0.5$ to $\sim 2\,$Gyr.
This implies that the star formation activity of the most massive
galaxies quickly declines for $z\lesssim 3$, i.e. that the redshift
distribution of SCUBA/MAMBO galaxies should peak at $z\gtrsim 3$, as
quasars reach their maximum luminosity (at $z\simeq 2.5$). This
explains why very luminous quasars are more easily detected at (sub)-mm
wavelengths for $z \gtrsim 2.5$.

A ``quasar phase'' follows, lasting $10^7$--$10^8\,$yrs, and a long
phase of passive evolution of galaxies ensues, with their colors
becoming rapidly very red [Extremely Red Object (ERO) phase].
Intermediate- and low-mass spheroids have lower Star Formation Rates
(SFRs) and less extreme optical depths. They show up as Lyman-Break
Galaxies (LBGs).

Therefore, in this scenario, large ellipticals evolve essentially as in
the ''monolithic'' scenario, yet in the framework of the standard
hierarchical clustering picture. Many aspects and implications of this
compound scheme have been addressed by our group in a series of papers
(Granato et al.\ 2001, Magliocchetti et al.\ 2001, Perrotta et al.\
2002, Romano et al.\ 2002). Here we only summarize how the scenario
compare with sub-mm counts.

\begin{figure*}                                                 
  \includegraphics[width=\textwidth,height=7cm]{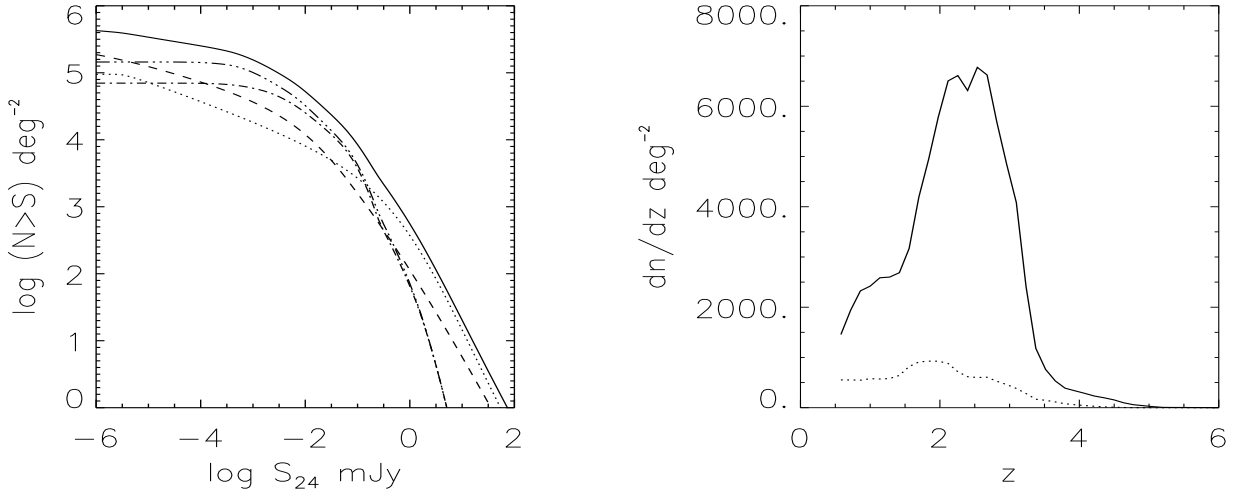} 
  \caption{Integral counts and $z$ distribution for a flux
  limit of $30\, \mu$Jy at $24\,\mu$m 
  by the model by Granato et al. (2001). Sybols for counts 
  as in Figure \ref{fig1}. The three-dots/dashed line shows the total counts of ellipticals,
  including also those where the star formation has ended. In the right-hand panel,
  the solid and the dotted lines show the $z$ distributions of ellipticals during
  and after the star formation phase, respectively.}  
  \label{fig2}                                                       
\end{figure*}                                                        

\subsection{Counts at (sub)-mm wavelengths}
The (sub)-mm counts are expected to be very steep because of the
combined effect of the strong cosmological evolution of dust emission in
spheroidal galaxies and of the strongly negative K-correction (the dust
emission spectrum steeply rises with increasing frequency). The model by
Granato et al. (2001) has extreme properties in this respect: above
several mJy its $850\,\mu$m counts reflect the high-mass exponential
decline of the mass function of dark halos. In this model, SCUBA/MAMBO
galaxies correspond to the phase when massive spheroids formed most of
their stars at $z\gtrsim 2.5$; such objects essentially disappear at lower
redshifts. On the contrary, the counts predicted by alternative models
(which are essentially phenomenological)
while steep, still have a power law
shape, and the redshift distribution has an extensive low-$z$ tail.
As illustrated by Figure~\ref{fig1}, the recent relatively large area surveys
are indeed suggestive of an
exponential decline of the $850\,\mu$m counts above several mJy.
Further evidence in this direction comes from MAMBO surveys at
$1.2\,$mm.
\adjustfinalcols

\subsection{Predictions for SIRTF surveys}

SIRTF surveys have the potential of providing further tests of the
model. In particular the $24\,\mu$m survey to be carried out as a
part of the GOODS (http://www.stsci.edu/science/goods) Legacy
Science project should reach a flux limit of $100\,\mu$Jy.
According to the model, about 50\% of detected galaxies should be
spheroidal galaxies forming their stars at $z \gtrsim 2$. About
400--600 such objects are expected over an area of 0.1 square
degree (Figure~\ref{fig2}). Their redshift distribution is predicted to
peak at $z$ slightly above 2, with a significant tail extending up
to $z\simeq 3$.


\end{document}